\pdfoutput=1
\documentclass[a4paper, 10pt, conference]{ieeeconf}      

\IEEEoverridecommandlockouts                              

\usepackage{graphics} 
\usepackage{epsfig} 
\usepackage{mathptmx} 
\usepackage{times} 
\usepackage{amsmath} 
\usepackage{amssymb}  

\usepackage{array}

\title{\LARGE \bf
Fault Tolerant Thermal Control of Steam Turbine Shell Deflections}

\author{Mert Geveci$^{1}$
\thanks{$^{1}$GE Global Research,
        1 Research Circle, Niskayuna, NY 12309, USA
        {\tt\small Mert.Geveci@ge.com}}%
}

\begin{document}

\maketitle
\thispagestyle{empty}
\pagestyle{empty}

\begin{abstract}
The metal-to-metal clearances of a steam turbine during full or part load operation are among the main drivers of efficiency. The requirement to add clearances is driven by a number of factors including the relative movements of the steam turbine shell and rotor during transient conditions such as startup and shutdown. This paper includes a description of a control algorithm to manage external heating blankets for the thermal control of the shell deflections during turbine shutdown. The proposed method is tolerant of changes in the heat loss characteristics of the system as well as simultaneous component failures.
\end{abstract}

\section{Introduction \label{sec:Problem-Description}}
Transient peak-to-peak shell deflections are the main driver in setting metal-to-metal clearances in steam turbines \cite{Ekbote2008}. These transient deflections are mainly caused by temperature differentials between the upper and lower halves of the steam turbine shell, which are in turn caused by the variations in heat loss characteristics of the two shell halves \cite{Ekbote2008} especially while the turbine shell cools down with the turbine shut down. 

\begin{figure}
\centering
\includegraphics[width=0.75\columnwidth]{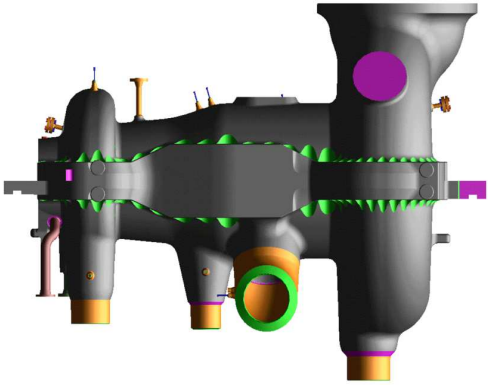}
\caption{\label{fig:Turbine-Outline} The combined high pressure/intermediate pressure (HP-IP) shell of a GE D11 steam turbine. }
\end{figure}

The combined high pressure/intermediate pressure (HP-IP) shell of a General Electric (GE) D11 steam turbine is shown in Figure \ref{fig:Turbine-Outline}. 
In this configuration, the inlets for both the HP and the IP turbine sections are in the middle bottom of the shell with the HP flow expanding towards the left and the IP flow expanding towards the right. All five inlet and outlet sections except one are located on the bottom half of the shell leading to a larger surface area. The larger surface area of the lower shell results in transient temperature differentials between the two shell halves, which in turn results in vertical deflections of the shell in \emph{U} and \emph{reverse-U} shapes.  

\begin{figure}
\centering{}\includegraphics[width=0.95\columnwidth]{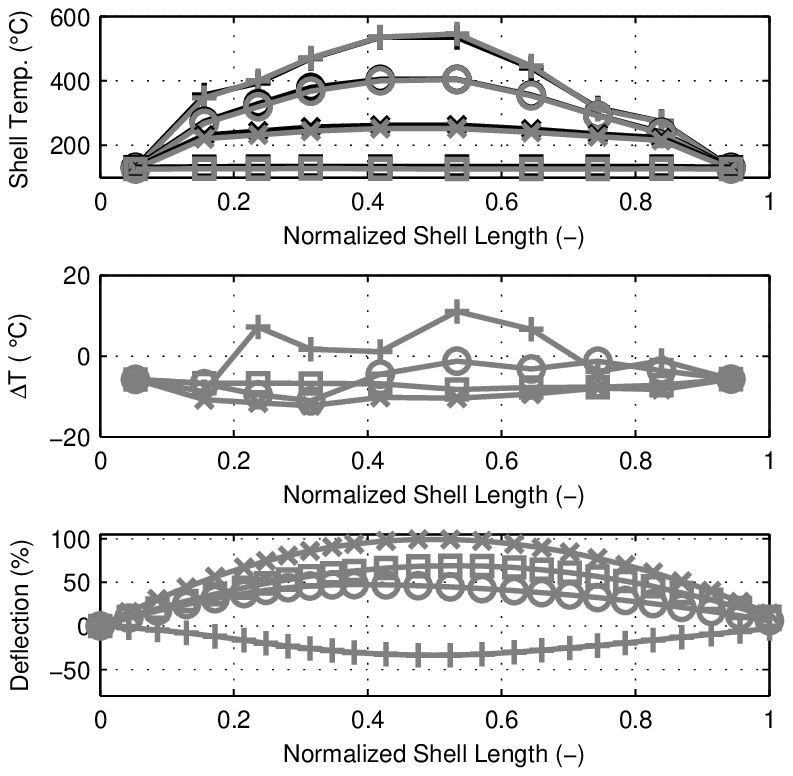}
\caption{\label{fig:Uncontrolled-results}The simulated time histories of individual shell temperatures, the temperature differentials, and the vertical deflections for the natural cooling case after a hot shutdown. In each figure, the markers "+", "$\circ$", "x", and "$\square$" represent t = 0, 37, 117, 298 hours, respectively. In the top figure, the dark lines represent the temperature levels of the upper shell while the lighter grey lines represent the temperature levels of the lower shell. Normalized length of unity represents the generator end of the turbine. In the bottom figure, the vertical deflections of the shell are normalized by the peak deflections.}
\end{figure}

A sample time history established by the simulation of a steam turbine shell model is shown in Figure \ref{fig:Uncontrolled-results}. 
In this case, time zero marks the time of turbine shutdown after hot operation and data is shown for over ten days of no turbine operation. For the vertical deflections shown here and the rest of study, the results are normalized by the peak deflection level observed during this uncontrolled cooldown event.

Various mechanical design features can be implemented to alleviate the vertical deflection issue. Most such approaches involve the redesign of the turbine shell and flow path to maintain uniform boundary conditions around the inner shell. Such design features are costly to implement and are not feasible for the improvement of the performance of in-service units, especially units with a single shell like the GE D11. A lower cost method to reduce the clearances of a steam turbine also has promise in the new unit space for designs with a single shell such as the GE D400. 

The installation of electric blankets on the outside of the steam turbine shell has been reported in many combined cycle and solar thermal units \cite{Spelling2011,Spelling2012}. The main purpose of these installations is to keep the turbine warm during shutdown periods, which enables faster plant startup, better management of low-cycle fatigue related life consumption, or a combination of the two. 

An additional potential use of the shell heating blankets is the control of the temperature differential between the two shell halves. Such active control would alleviate the peak levels of shell deflections. Once lower peak shell deflections are established, the metal-to-metal clearances can be reduced by modification of the packings installed inside the shell. Such a modification may result in significant improvements in the hot operation efficiency of the steam turbine.

The main challenge associated with the reduction of clearances by active thermal control of shell deflections is ensuring the high reliability required. Failure to adequately manage the temperature differential between the two shell halves may result in rubbing of the rotor against the shell and a permanent loss of efficiency as well as other operational issues such as vibration problems. This is in contrast with clearance control systems that are based on cooling stator parts to shrink them - typically applied in gas turbines \cite{Korson1995}. Such systems are fail-safe in nature and the loss of clearance control function (\emph{i.e.,} cooling) results in an increase in the level of available clearances (and temporary loss of efficiency) rather than movement of the parts towards each other risking interference. In the steam turbine case, the control system is required to avoid interference after the clearances have been reduced by hardware changes.  The control system is then required to have reliability comparable to the major turbine components it is protecting. The replacement of individual heating system components should be limited to scheduled outages to the extent possible since the heating blankets and the associated instrumentation are installed under turbine insulation. All of these factors combined require the design of a system that is tolerant of

\begin{itemize}
\item Simultaneous failures of multiple heaters;
\item Failures of individual temperature sensors; and
\item Variations in heat transfer characteristics caused by initial installation effects or disturbances associated with maintenance activities.
\end{itemize}

The requirements related to the temperature sensors can be addressed through physical or analytical redundancy and are beyond the scope of this study. The failures of heaters and changes in the heat transfer characteristics of the system, however, are within scope. 

The original requirement of keeping the steam turbine shell warm during shutdown periods \cite{Spelling2012} is still valid, but is secondary to the deflection control requirements as the controlled cooling of the turbine shell only results in longer start times, while the failure to control differential temperature between the shells results in damage to the turbine. Therefore, the robustness requirements come with the option to trade the average temperature of the shell against the temperature differential between the shell halves. 

The main focus of this study is the design of a controller that is capable of managing both the average temperature of the steam turbine shell and the vertical deflections under nominal conditions and the robust control of the vertical deflections under off-nominal conditions.

\section{System Architecture\label{sec:System-Architecture}}
\begin{figure}
\centering{}\includegraphics[width=0.9\columnwidth]{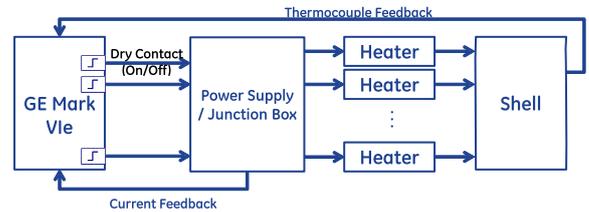}
\caption{\label{fig:Hardware-Architecture}A high level overview of the control system hardware architecture}
\end{figure}

A high level overview of the control system hardware architecture is shown in Figure \ref{fig:Hardware-Architecture}. 
The application code is implemented in a GE Mark VIe\footnote{Trademark of General Electric Company} controller with the primary output of the controller in the form of dry contacts that enable the closing of the circuits for the individual heaters. The analog inputs from the system include thermocouple inputs and current feedback measurements from each \emph{heating zone}. In the context of this manuscript, a \emph{heating zone} is defined as an individually controlled heating blanket coupled with a thermocouple feedback representing the average temperature of the particular zone.

The details of the selection of the number of heating zones and their placement are beyond the scope of this manuscript as this process is highly dependent on the geometry of the turbine shell of interest as well installation related constraints such as the ease of installation and maintainability. The example case considered here is representative of a GE D11 steam turbine and includes 20 individual heating zones. 

\section{Process Model\label{sec:Process-Model}}

A process model has been developed both for algorithm validation and potentially for embedded implementation. The deflection sub-model is implemented as a part of the control system. Furthermore, the model was developed with an emphasis on computational efficiency and is not suited for the detailed prediction of turbine performance during and after shutdown. A higher order, multi-directional model such as the one described by \cite{Spelling2012a} would be required for such a multi-faceted analysis. 

For the modeling of the vertical deflections, the thermal dynamics of the shell are assumed to be independent of shell deflections. It is further assumed that all non-thermal conditions are constant from the perspective of shell deflections. In other words, only the portion of shell deflections that are driven by shell temperature changes are modeled. The impact of factors such as gravity and existing deflections of the shell are not considered. It is common practice to compensate for these permanent effects by adjusting the natural shape of the rotor, which mitigates the need for active control.

\subsection{Thermal Model of the Steam Turbine Shell\label{sub:Thermal-Model}}

\begin{figure}
\centering{}\includegraphics[width=0.95\columnwidth]{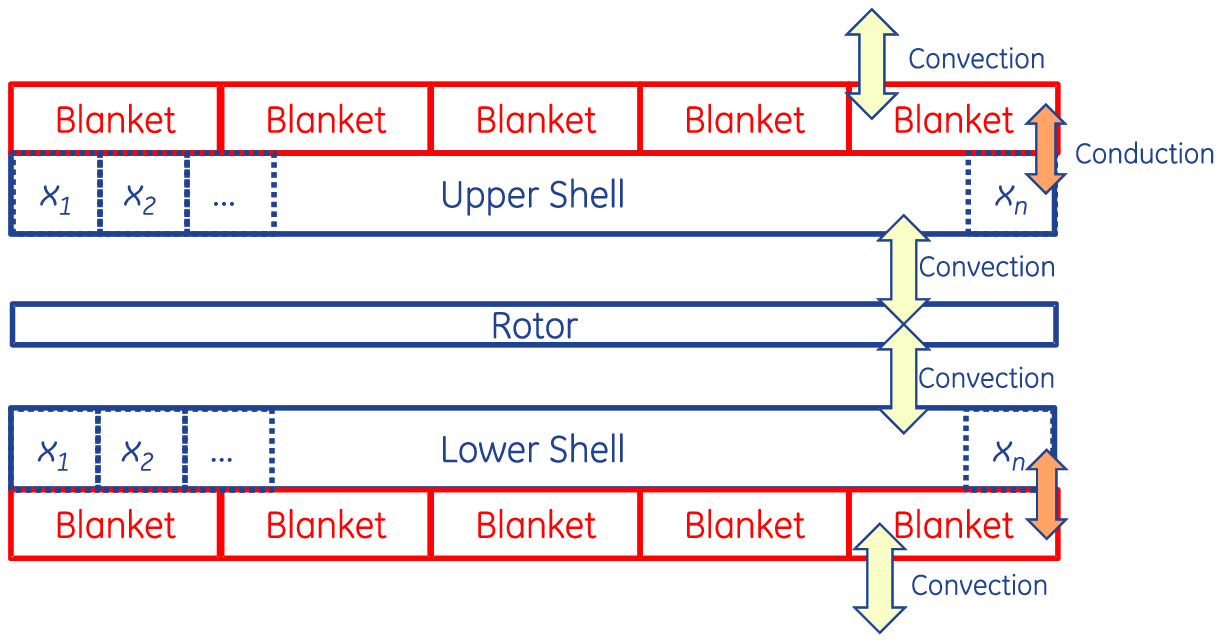}\protect\caption{\label{fig:Thermal-model-structure} Model structure for the thermal
shell model}
\end{figure}

The structure of the thermal model of the steam turbine shell is shown
in Figure \ref{fig:Thermal-model-structure}.
In this structure, the shell itself is represented by a finite difference
model with a total of $N$ elements. For the purposes of this study, each shell half was discretized into a total of 10 elements and each shell element was assumed to have a corresponding heating zone associated with it.

The heating blankets can be modeled as lumped thermal masses with
electrical heaters embedded in them. The generation of heat is assumed to be instantaneous
where the state of each heater is binary (\emph{i.e.,} each heater
is on or off at each time step and is not continuously modulated).
It is assumed that heating blankets transfer heat to the environment
by convection and to the shell by conduction. Heat transfer among
blankets is ignored since the heating elements are typically physically
separated from each other to avoid local heating issues and are embedded
in enclosures that are poor thermal conductors. The heat transfer
from the blankets to the shell is modeled utilizing the contact resistance
concept \cite{Holman1992} in the form
\begin{eqnarray*}
q & = & A_{s}\frac{T_{A}-T_{B}}{1/h_{c}}
\end{eqnarray*}
where $T_{A}$ and $T_{B}$ represent the two surfaces in contact;
$h_{c}$ is the contact coefficient; $A_{s}$ is the contact surface area.
The determination of the contact resistance for various engineering
materials and varying surface condition is often difficult and is
the subject of many studies \cite{Madhusudana1986}. For the case
of heating blankets installed on a steam turbine the problem of determining
the contact resistance is especially challenging since the contact resistance
can significantly vary based on the care applied during the installation
of the blankets. Careful placement and attachment of each blanket to
the shell will result in a smaller air gap and low contact resistance,
however, such an installation is not always guaranteed due to operation
constraints associated with installing on a turbine shell. Therefore, it is important that the overall control system be relatively insensitive to these installation
variations. 

The rotor can be modeled in a lumped parameter fashion as a single node.
It is assumed that the rotor can transfer heat to each shell element
via convection \cite{Marinescu2015a}. The prediction of the rotor
temperature is critical for the proper budgeting and management of
the low cycle fatigue impact \cite{Marinescu2012}. However, since
the primary interest of this document is the modeling of temperature
variations across the shell, the impact of the inaccuracies in capturing
shell-to-rotor heat transfer characteristics is less critical. 
The potential impact of the variability in the shell-to-rotor heat
transfer characteristics is captured by assigning a large variability
band to the corresponding heat transfer coefficient as described in
Section \ref{sec:Fault-Tolerance-Simulation-Resul}.

The thermal dynamics of the shell were modeled using the
heat equation which can be written in its one-dimensional form as

\[
\frac{1}{\alpha}\frac{\partial T}{\partial t}=\frac{\partial^{2}T}{\partial x^{2}}+\frac{\dot{q}}{k}
\]
where temperature is represented by $T$; time is represented by $t$;
the heat input per unit volume is represented by $\dot{q}$; and the
thermal diffusivity and thermal conductivity of the material are represented
by $\alpha$ and $k$, respectively \cite{Incropera1990}. 

For a numerical solution, the partial differential equation can be further
reduced to ordinary differential equations using a first-order discrete
representation \cite{LeVeque2007} of the spatial derivative as

\[
\frac{\partial^{2}T}{\partial x^{2}}=\frac{1}{h^{2}}\left[T\left(x-h\right)-2T\left(x\right)+T\left(x+h\right)\right]
\]
which is valid only for internal elements . Assuming Neumann
boundary conditions \cite{LeVeque2007}, one can obtain the simplified set of ODEs equations
in the form

\[
\frac{1}{h^{2}}\left[\begin{array}{cccccc}
-h & h\\
1 & -2 & 1\\
 &  & \ddots\\
 &  &  & 1 & -2 & 1\\
 &  &  &  & h & -h
\end{array}\right]\left[\begin{array}{c}
T_{0}\\
T_{1}\\
T_{2}\\
T_{3}\\
\vdots\\
T_{m}\\
T_{m+1}
\end{array}\right]\cdots
\]

\begin{eqnarray*}
+\left[\begin{array}{c}
0\\
\frac{\dot{q}_{1}}{k}\\
\frac{\dot{q}_{2}}{k}\\
\frac{\dot{q}_{3}}{k}\\
\vdots\\
\frac{\dot{q}_{m}}{k}\\
0
\end{array}\right] & = & \left[\begin{array}{c}
\frac{h}{2}\frac{1}{\alpha}\frac{\partial T_{0}}{\partial t}\\
\frac{1}{\alpha}\frac{\partial T_{1}}{\partial t}\\
\frac{1}{\alpha}\frac{\partial T_{2}}{\partial t}\\
\frac{1}{\alpha}\frac{\partial T_{3}}{\partial t}\\
\vdots\\
\frac{1}{\alpha}\frac{\partial T_{m}}{\partial t}\\
\frac{h}{2}\frac{1}{\alpha}\frac{\partial T_{m+1}}{\partial t}
\end{array}\right]
\end{eqnarray*}
The first and the last elements of the
input vector are equal to zero since there are no heating blankets
corresponding to the ends of the turbine. These areas are generally
kept free of intrusions for ease of service access. 

In this representation, the inputs $\dot{q}_{m}$ are each a sum of all \emph{external} heat that is being transferred to a particular shell element, which
includes the heat transfer from the blankets as well as the heat transfer
from the shell to the rotor. The dynamics of the blankets and the
rotor are handled outside the scope of the shell model.  Each of these elements are represented by one corresponding additional state.

\subsection{Steam Turbine Shell Deflection Model\label{sub:Shell-Deflection-Model}}
A model linking the temperature differential between the two shell halves to the vertical deflection of the shell is required in order to quantify the potential for turbine clearance reduction. As discussed above, such a model is primarily intended for validation purposes and for model-based control purposes. Both the brute force robustness validation process described in Section \ref{sec:Fault-Tolerance-Simulation-Resul} and real-time control require a computationally efficient and robust model that can be executed at a speed significantly faster than real-time.

The thermal model described in Section \ref{sub:Thermal-Model} corresponds
to a beam from a mechanics perspective and can be modeled as such.
In order to model the impact of the temperature differential between
the halves of the beam, one can represent the system by a beam with a
temperature variation from top to bottom. One such model is provided
by \cite{Roark2002} for a beam with a uniform temperature variation. The configuration
of interest here, however, involves a beam with a horizontally non-uniform (and time-varying) temperature profile
across the length of the beam. Again, in order to fit the discrete
nature of the finite difference model described in Section \ref{sub:Thermal-Model},
one can assume that the shell is a combination of stringed beams with
a uniform temperature differential across (\emph{i.e.,} top to bottom)
each. The unknown boundary conditions for the
joining points can be determined by matching the deflection
and slope at each joining point. Even though a numerical solution
would be feasible, an analytical solution is preferred
due to the computational requirements. The determination of such an analytical solution is shown
here for the case of two beam elements for simplicity. The approach
can be readily extended to a beam with a larger number of temperature
zones. Simulation results shown elsewhere in this study were obtained with five beam elements.

Consider a two element beam configuration with the two beams joined in the middle and simply supported on each end. In this configuration, $\theta_{A}$, $\theta_{B}$, and $\theta_{C}$ represent the slopes at the left end, joining point, and the right end: $A$, $B$, and $C$, respectively. $\Delta T_{1}$ and $\Delta T_{2}$ represent the top-to-bottom temperature differentials for the beams $1$ and $2$. The vertical deflections at points $A$, $B$, and
$C$ are represented by $y_{A}$, $y_{B}$, and $y_{C}$. The parameters of the model are the temperature coefficient of expansion and the depth of the beam which are represented by $\gamma$ and $t_{b}$, respectively.
Finally, the horizontal position along the beam is represented by the independent variables $x_{1}$ and $x_{2},$ where $x_{1}$ is the horizontal distance from point $A$ and $x_{2}$ is the horizontal distance from point $B$. The lengths of the two elements are represented by $l_{1}$ and $l_{2}$. Utilizing Roark's deflection formulas for
a beam with uniform temperature variation, the slope and deflection can be written for each element as

\begin{eqnarray*}
\theta_{1}=\theta_{A}+\frac{\gamma}{t_{b}}\Delta T_{1}x_{1}; &  & \theta_{2}=\theta_{C}+\frac{\gamma}{t_{b}}\Delta T_{2}x_{2}
\end{eqnarray*}

\begin{eqnarray*}
y_{1}=\theta_{A}x_{1}+\frac{\gamma}{2t_{b}}\Delta T_{1}x_{1}^{2}; &  & y_{2}=\theta_{C}x_{2}+\frac{\gamma}{2t_{b}}\Delta T_{2}x_{2}^{2}
\end{eqnarray*}
Substituting $\theta_{2}=-\theta_{1}$ and matching the boundary conditions at point $B$

\begin{equation*}
l_{1}\theta_{A}-l_{2}\theta_{C}=\frac{\gamma}{2t_{b}}\left(\Delta T_{2}l_{2}^{2}-\Delta T_{1}l_{1}^{2}\right)
\end{equation*}

\begin{equation*}
\theta_{A}+\theta_{C}=-\frac{\gamma}{t_{b}}\left(\Delta T_{1}l_{1}+\Delta T_{2}l_{2}\right)
\end{equation*}
Solving for $\theta_{A}$ and $\theta_{C}$ one obtains

\begin{eqnarray*}
\theta_{C} & = & \frac{\frac{\gamma}{2l_{1}t_{b}}\left(\Delta T_{2}^{2}l_{2}^{2}-\Delta T_{1}^{2}l_{1}^{2}\right)-\frac{\gamma}{t_{b}}\left(\Delta T_{1}l_{1}+\Delta T_{2}l_{2}\right)}{\frac{l_{2}}{l_{1}}+1}
\end{eqnarray*}

\begin{eqnarray*}
\theta_{A} & = & \frac{l_{2}}{l_{1}}\theta_{C}+\frac{\gamma}{2l_{1}t_{b}}\left(\Delta T_{2}l_{2}^{2}-\Delta T_{1}l_{1}^{2}\right)
\end{eqnarray*}

\section{Control Concept Description\label{sec:Concept-Descriptions}}
\begin{figure}
\centering{}\includegraphics[width=0.95\columnwidth]{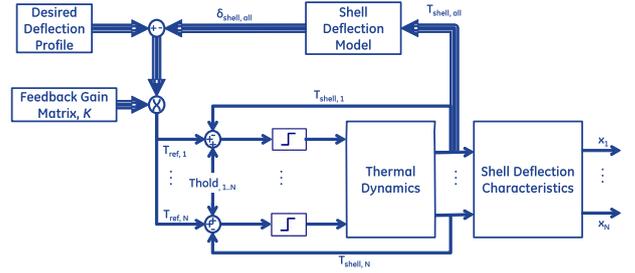}
\caption{\label{fig:Controller-Block-Diagram} Block diagram of the feedback control system }
\end{figure}

A block diagram of the active clearance control system is shown in Figure \ref{fig:Controller-Block-Diagram}. 
The design of the controller reflects the dual control goals of maintaining shell deflections and the overall average temperature of the shell simultaneously. The inner loop of the controller consists of individual controllers for each shell heating zone with a fixed gain relay for each loop. The error calculation and the switching decision for the relay are executed within the microprocessor based controller, while the actual switching is realized through a dry contact relay. 

The outer loop of the controller utilizes the deflection model described in Section \ref{sub:Shell-Deflection-Model} to estimate the current deflection profile of the shell. The estimated deflection values are individually compared against a desired deflection profile. The calculated error value is then multiplied by a square feedback gain matrix, $K$. For the purposes of the work summarized here, the feedback gain matrix, $K$, was configured to be a scalar matrix with all of the diagonal values equal to $50$, which was selected based on observed interactions of the outer loop with the inner loop as the oscillation frequency of the inner loop would be affected by outer loop action. Such interactions are not desirable as more frequent cycling of the heaters is likely to result in shorter relay life. The selection of both the diagonal and off-diagonal terms of the feedback gain matrix are subjects for further study, as populating the off-diagonal terms could further improve the fault tolerance properties of the closed loop system by utilizing blankets across the shell to compensate for individual failures. 

The feedback signals produced by the deflection feedback loop are then summed with the individual hold temperature references. These temperature references are determined based on the \emph{natural} axial temperature profile of the shell and a desired minimum hold temperature selected based on considerations of rotor life, start time, and heating system life considerations. 

\begin{figure}
\centering{}\includegraphics[scale=0.95]{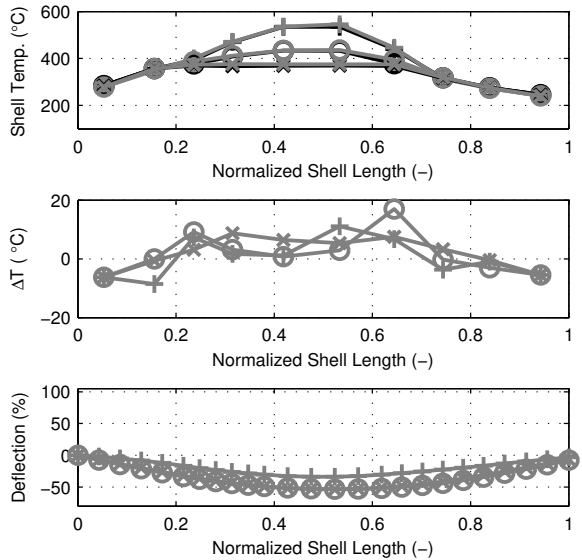}
\caption{\label{fig:Controlled-results}The simulated time histories of individual shell temperatures, the temperature differentials, and the vertical deflections for the controlled cooling case after a hot shutdown. In each figure, the markers "+", "$\circ$", and "x" represent t = 0, 37, and 117 hours, respectively. In the top figure, the dark lines represent the temperature levels of the upper shell while the lighter grey lines represent the temperature levels of the lower shell. Normalized length of unity represents the generator end of the turbine.}
\end{figure}

A plot showing a typical controlled cooldown is shown in Figure \ref{fig:Controlled-results}. 
In this case, the temperature levels of the individual heating zones are controlled to the temperature references determined by the desired hold temperature levels and the feedback from the deflection loop. 

\section{Fault-Tolerance Simulation Results\label{sec:Fault-Tolerance-Simulation-Resul}}
\begin{figure}
\includegraphics[width=0.9\columnwidth]{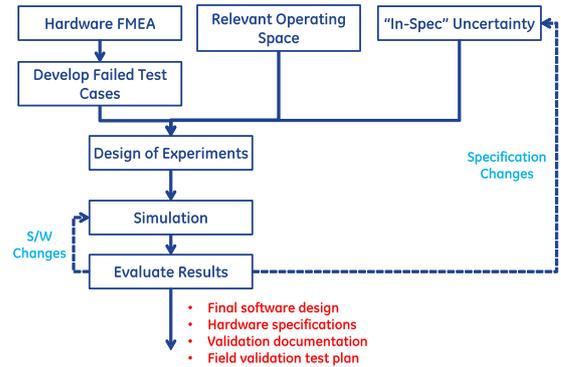}
\caption{\label{fig:Robustness-testing-process} Process utilized to ensure the robustness of the control system}
\end{figure}

The model described in Section \ref{sec:Process-Model} was tuned to match a combination of measured data and a high fidelity finite elements model. The tuned model was then utilized to assess the robustness of the control system described in Section \ref{sec:Concept-Descriptions}. The robustness of the closed loop system was assessed utilizing the process shown in Figure \ref{fig:Robustness-testing-process} where the impact of both nominal variability and failure modes are considered.

\begin{figure}
\centering{}\includegraphics[scale=0.95]{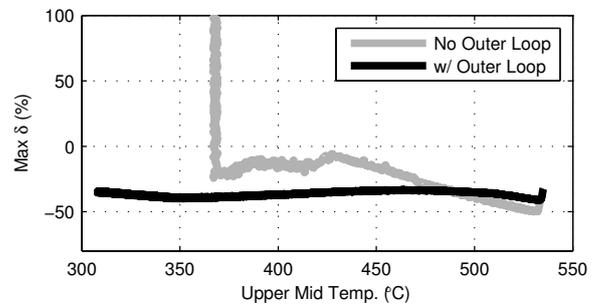}
\caption{\label{fig:Two-failed} A comparison of the response of the inner-loop-only controller vs. the controller with the outer loop active. Failures of two middle heaters on the lower shell were simulated.}
\end{figure}

In Figure \ref{fig:Two-failed}, a comparison of the response of only the inner loop of the controller (with a fixed temperature differential offset added to make the nominal responses equivalent) against the response of the overall control system is shown for the case of two failed middle blankets on the upper shell. 
Until the temperature level reaches the hold point, the responses are not substantially different, while the response during the \emph{hold} portion of the cooldown results in significant positive deflections (over 200\%, not fully shown) for the inner loop controller. In fact, the response of the inner loop controller is significantly worse than the case with no heating system as shown in Figure \ref{fig:Uncontrolled-results}. The outer loop, on the other hand, is able to mitigate the impact of the failed blankets in the lower shell by lowering the temperature references for the corresponding upper shell blankets.

\begin{figure}
\centering{}\includegraphics[width=0.95\columnwidth]{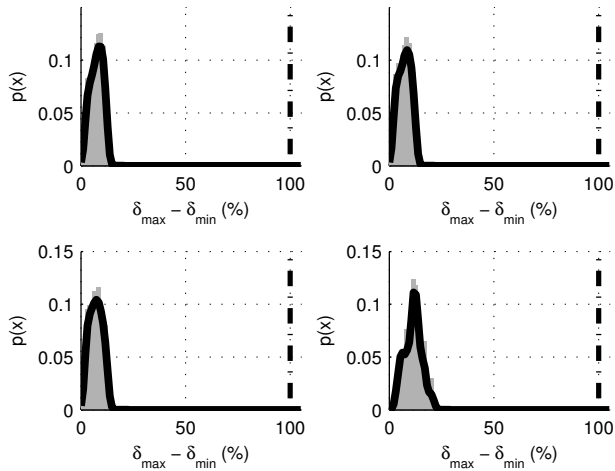}
\caption{\label{fig:Robustness-results} The probability density function estimates resulting from the fault-tolerance simulations results. The sub-plots left-to-right and top-to-bottom represent: one heater failed at a time, two, three, and four heaters failed simultaneously. All possible failed blanket combinations were simulated.}
\end{figure}

Following the process described in Figure \ref{fig:Robustness-testing-process}, hardware failure modes were simulated along with the in-process variability associated with each subsystem. In order to represent component variability, each heat transfer coefficient was assumed to be uniformly distributed within 25\% of its nominal value with the exception of contact resistance between the shell and the heaters. Due to potential installation variability, the contact resistance was assumed to be only within 50\% of its nominal value. The estimates of the probability density functions associated with varying levels of heater failure combinations are shown in Figure \ref{fig:Robustness-results}. 
For comparison purposes, the reference shell movement can be adopted from Figure \ref{fig:Uncontrolled-results} where the peak-to-peak deflections of the middle point of the shell were 133\%. The reductions in peak-to-peak shell deflections are kept above 80\% for all cases with one to four heating zone failures. It must also be noted that the peak-to-peak deflections of 133\% were for the case of a passive cooldown with turbine insulation performing exactly to the specification. In practice, the peak-to-peak deflection may need to be selected two to three times larger than this best case scenario condition since insulation installation variations often cause increased shell deflections. 

In addition to the cases shown in Figure \ref{fig:Robustness-results}, further failure modes were evaluated and the results are summarized in Table \ref{tab:Robustness-results}. 
\begin{table}
\caption{\label{tab:Robustness-results}Summary of fault tolerance simulation results}
\centering{}\begin{tabular}{| >{\centering}m{2.5cm} | >{\centering}m{1.5cm} | >{\centering}m{3cm} |}
\hline 
Case & Peak-to-Peak Deflection (\%) & Mitigation\tabularnewline
\hline 
\hline 
One to four failed heaters & 22 & None required\tabularnewline
\hline 
Five failed heaters & 29 & None required\tabularnewline
\hline 
One heater stuck On & 22 & None required\tabularnewline
\hline 
Two heaters stuck & 30 & Maintenance procedure\tabularnewline
\hline 
Out-of-spec air gaps (2) & 14 & None required\tabularnewline
\hline 
6x heat losses - lower shell & 34 & None required\tabularnewline
\hline 
7x heat losses - lower shell & 68 & Implement insulation specification\tabularnewline
\hline 
\end{tabular}
\end{table}
Simulation of five failed heaters further showed the robustness of the system to actuator losses. Cases with heaters stuck resulted in increases in deflection larger compared to the loss of heaters. One mitigating factor here is that a stuck heater is expected to be caused by relay failure. Relays are generally located inside the control cabinet and are easier to access compared to components mounted on the shell. The sensitivity to out-of-spec air gaps (simulated as a factor of 20 increase in contact resistance) is also fairly low. Increased heat losses (only on one side of the turbine) can be managed up to a factor of six increase. When the heat losses are increased beyond this point, the heaters are basically not able to keep up - even with the corresponding heaters on the other half of the turbine inactive. This level of heat loss variability is uncommon as it would result in much larger vertical deflections on a shell with no heaters.

\section{Conclusions\label{sec:Conclusions}}
A model-based control approach to thermally control the deflections of a steam turbine shell has been presented. The resulting controller has been demonstrated in simulation to be tolerant of a large number of failure modes and variation in the heat transfer characteristics of the system. 

The proposed robustness demonstration process is brute force in nature, but is highly effective in determining the capability limitations of a given control system. The results can be utilized to further improve the fault tolerance properties of the control system as well as to improve the design of the physical system.

The capabilities of the thermal shell deflection control system could be further improved by studying one of several areas. One obvious area of study is the development of a partially or fully populated feedback gain matrix. Such an approach has the potential to further improve the robustness and the transient capability of the system by compensating for deflection tracking errors utilizing the highly coupled nature of the plant. An additional area of study is the controllability and observability properties of the system. A study of the controllability properties of the system may lead to formal methods for the placement of individual heating zones. A study of the observability properties of the system may lead to a reduction in the number of sensors in the system by determining a minimum viable sensor set.  

\addtolength{\textheight}{-12cm}   




\bibliographystyle{unsrt}
\bibliography{Bibliography,./Geveci_bib}

\begin{thebibliography}{10}

\bibitem{Ekbote2008}
Ajit~B. Ekbote and Howard~M. Brilliant.
\newblock The analytical approach to the temperature prediction of steam
  turbine shells based on the thermocouple temperature measurements at a few
  points.
\newblock In {\em Structures and Dynamics, Part A}, volume~5, pages 59 -- 68,
  Berlin, Germany, 2008.

\bibitem{Spelling2011}
James Spelling, Markus Jocker, and Andrew Martin.
\newblock Thermal modeling of a solar steam turbine with a focus on start-up
  time reduction.
\newblock In {\em Proceedings of the ASME Turbo Expo 2011}, volume~3, pages
  1021 -- 1030, Vancouver, BC, Canada, 2011.

\bibitem{Spelling2012}
James Spelling, Markus Jocker, and Andrew Martin.
\newblock Annual performance improvement for solar steam turbines through the
  use of temperature-maintaining modifications.
\newblock {\em Solar Energy}, 86(1):496 -- 504, 2012.

\bibitem{Korson1995}
Shannon Korson and Arthur~J. Helmicki.
\newblock ${H}_{\infty}$ based controller for a gas turbine clearance control
  system.
\newblock In {\em Proceedings of the 4th IEEE Conference on Control
  Applications}, pages 1154 -- 1159, 1995.

\bibitem{Spelling2012a}
James Spelling, Markus Jocker, and Andrew Martin.
\newblock Thermal modeling of a solar steam turbine with a focus on start-up
  time reduction.
\newblock {\em Journal of Engineering for Gas Turbines and Power}, 134(1),
  2012.

\bibitem{Holman1992}
Jack~P. Holman.
\newblock {\em Heat Transfer}.
\newblock McGraw-Hill, {Seventh} edition, 1992.

\bibitem{Madhusudana1986}
C.V. Madhusudana and Leroy~S. Fletcher.
\newblock Contact heat transfer - the last decade.
\newblock {\em AIAA journal}, 24(3):510 -- 523, 1986.

\bibitem{Marinescu2015a}
Gabriel Marinescu, Peter Stein, and Michael Sell.
\newblock Natural cooling and startup of steam turbines: Validity of the
  over-conductivity function.
\newblock {\em Journal of Engineering for Gas Turbines and Power}, 137(11),
  2015.

\bibitem{Marinescu2012}
Gabriel Marinescu and Andreas Ehrsam.
\newblock Experimental investigation into thermal behavior of steam turbine
  components. {P}art 2 - natural cooling of steam turbines and the impact on
  {LCF} life.
\newblock volume~4, pages 1111 -- 1120, Copenhagen, Denmark, 2012.

\bibitem{Incropera1990}
Frank~P. Incropera and David P.~De Witt.
\newblock {\em Fundamentals of Heat and Mass Transfer}.
\newblock Wiley \& Sons, 1990.

\bibitem{LeVeque2007}
Randall~J. Le{V}eque.
\newblock {\em Finite Difference Methods for Ordinary and Partial Differential
  Equations}.
\newblock Society for Industrial and Applied Mathematics (SIAM), 2007.

\bibitem{Roark2002}
Warren~C. Young and Richard~G. Budynas.
\newblock {\em Roark''s Formulas for Stress and Strain}.
\newblock McGraw-Hill, seventh edition, 2002.

\end{thebibliography}

\end{document}